\documentclass[10pt,twoside]{cpc-hepnp}

\usepackage{multicol}
\usepackage{graphicx}
\usepackage{amssymb,bm,mathrsfs,bbm,amscd}
\usepackage[tbtags]{amsmath}
\usepackage{lastpage}

\def\kbar{\overline{K}{}^{\,0}}
\def\dbar{\overline{D}{}^{\,0}}

\def\cp{$CP$}
\def\cpv{$CPV$}
\def\ra{\!\rightarrow\!}

\def\dklnu{$D^0\ra K^+\ell^-\nu$}
\def\dkpi{$D^0\ra K^+\pi^-$}

\def\dkkpp{$D^0\ra K^+K^-/\pi^+\pi^-$}
\def\dkspp{$D^0\ra K^0_S\,\pi^+\pi^-$}
\def\ycp{$y^{}_{\rm CP}$}

\def\meve{~MeV}

\def\babar{Babar}

\def\simge{\mathrel{%
   \rlap{\raise 0.511ex \hbox{$>$}}{\lower 0.511ex \hbox{$\sim$}}}}
\def\simle{\mathrel{
   \rlap{\raise 0.511ex \hbox{$<$}}{\lower 0.511ex \hbox{$\sim$}}}}

\begin{document}

\title{\boldmath{Measurements of $D^0$-$\dbar$ Mixing and
Searches for \cp\ Violation: HFAG Combination of all Data}}

\author{%
A. J. Schwartz$^{1;1)}$\email{schwartz@physics.uc.edu} \\
(representing the HFAG charm group)
}

\maketitle

\address{
1~(Physics Department, University of Cincinnati, Cincinnati, Ohio 45221, USA)\\
}

\begin{abstract}
We present world average values for $D^0$-$\dbar$ mixing parameters
$x$ and $y$, \cp\ violation parameters $|q/p|$ and ${\rm Arg}(q/p)$,
and strong phase differences $\delta$ and $\delta^{}_{K\pi\pi}$. 
These values are calculated by the Heavy Flavor Averaging Group (HFAG) 
by performing a global fit to relevant experimental measurements. The results 
for $x$ and $y$ differ significantly from zero and are inconsistent 
with no mixing at the level of~$6.7\sigma$.
The results for $|q/p|$ and ${\rm Arg}(q/p)$ are consistent 
with no \cp\ violation. The strong phase difference $\delta$ is
less than $45^\circ$ at 95\%~C.L.
\end{abstract}



\footnotetext[0]{\hspace*{-2em}\small\centerline{\thepage\ --- \pageref{LastPage}}}%

\begin{multicols}{2}

\section{Introduction}

Mixing in the $D^0$-$\dbar$ system has been searched for for more than 
two decades without success --- until last year. Three experiments
--\,Belle,\cite{belle_kk} Babar,\cite{babar_kpi} and CDF\cite{cdf_kpi}\,-- 
have now observed evidence for this phenomenon. These measurements can 
be combined with others to yield World Average (WA) values for the 
mixing parameters 
$x\equiv(m^{}_1-m^{}_2)/\Gamma$ and 
$y\equiv\Delta (\Gamma^{}_1-\Gamma^{}_2)/(2\Gamma)$, where 
$m^{}_1,\,m^{}_2$ and $\Gamma^{}_1,\,\Gamma^{}_2$ are
the masses and decay widths for the mass eigenstates
$D^{}_1\equiv p|D^0\rangle-q|\dbar\rangle$ and
$D^{}_2\equiv p|D^0\rangle+q|\dbar\rangle$,
and $\Gamma=(\Gamma^{}_1+\Gamma^{}_2)/2$. 
Here we use the phase convention $CP|D^0\rangle=-|\dbar\rangle$
and $CP|\dbar\rangle=-|D^0\rangle$.
In the absence of \cp\ violation (\cpv), $p=q=1/\sqrt{2}$ and
$D^{}_1$ is \cp-even, $D^{}_2$ is \cp-odd.

Such WA values have been calculated by the Heavy Flavor Averaging 
Group (HFAG)\cite{hfag_charm} in two ways:
{\it (a)}\ adding together three-dimensional log-likelihood functions 
obtained from various measurements for parameters $x$, $y$, and $\delta$, 
where $\delta$ is the strong phase difference between amplitudes 
${\cal A}(D^0\ra K^+\pi^-)$ and ${\cal A}(D^0\ra K^-\pi^+)$;
and {\it (b)}\ doing a global fit to measured observables for 
$x$, $y$, $\delta$, an additional strong phase $\delta^{}_{K\pi\pi}$, and 
$R^{}_D\equiv\left|{\cal A}(D^0\ra K^+\pi^-)/{\cal A}(D^0\ra K^-\pi^+)\right|^2$. 
For this fit, correlations among observables are accounted for by using 
covariance matrices provided by the experimental collaborations. 
The first method has the advantage that non-Gaussian 
errors are accounted for, whereas the second method has the 
advantage that it is easily expanded to allow for \cpv. In 
this case three additional parameters are included in the fit:
$|q/p|$, $\phi\equiv {\rm Arg}(q/p)$, and
$A^{}_D\equiv (R^+_D-R^-_D)/(R^+_D+R^-_D)$, where the $+\,(-)$
superscript corresponds to $D^0\,(\dbar)$ decays. When both methods 
are applied to the same set of observables, almost identical results 
are obtained. The observables used are from measurements of \dklnu, 
\dkkpp, \dkpi, $D^0\ra K^+\pi^-\pi^0$, $D^0\ra K^+\pi^-\pi^+\pi^-$, 
and \dkspp\ decays, and from double-tagged branching 
fractions measured at the $\psi(3770)$ resonance.

Mixing in heavy flavor systems such as that of $B^0$ and $B^0_s$ 
is governed by the short-distance box diagram. In the $D^0$ system,
however, this diagram is doubly-Cabibbo-suppressed relative to 
amplitudes dominating the decay width, and it is also GIM-suppressed.
Thus the short-distance mixing rate is tiny, and $D^0$-$\dbar$ 
mixing is expected to be dominated by long-distance 
processes. These are difficult to calculate reliably, and 
theoretical estimates for $x$ and $y$ range over two-three 
orders of magnitude.\cite{BigiUraltsev,Petrov}

With the exception of $\psi(3770)\ra DD$ measurements, all methods 
identify the flavor of the $D^0$ or $\dbar$ when produced by 
reconstructing the decay $D^{*+}\ra D^0\pi^+$ or $D^{*-}\ra\dbar\pi^-$; 
the charge of the accompanying pion identifies the $D$ flavor. For signal 
decays, $M^{}_{D^*}-M^{}_{D^0}-M^{}_{\pi^+}\equiv Q\approx 6$\meve, 
which is relatively close to the threshold. Thus analyses typically
require that the reconstructed $Q$ be small to suppress backgrounds. 
For time-dependent measurements, the $D^0$ decay time is 
calculated via $(\ell/p)\times M^{}_{D^0}$, where $\ell$ is
the distance between the $D^*$ and $D^0$ decay vertices and 
$p$ is the $D^0$ momentum. The $D^*$ vertex position is 
taken to be at the primary vertex\cite{cdf_kpi} ($\bar{p}p$)
or is calculated from the intersection of the $D^0$ momentum 
vector with the beamspot profile ($e^+e^-$).

\section{\boldmath Input Observables}

The global fit determines central values and errors for
eight underlying parameters using a $\chi^2$ statistic
constructed from 26 observables. The
underlying parameters are $x,\,y,\,\delta,\,R^{}_D,
A^{}_D,\,|q/p|,\,\phi$, and $\delta^{}_{K\pi\pi}$.
The parameters $x$ and $y$ govern mixing, and the
parameters $A^{}_D$, $|q/p|$, and $\phi$ govern \cpv.
The parameter $\delta^{}_{K\pi\pi}$ is the strong phase 
difference between the amplitude ${\cal A}(D^0\ra K^+\pi^-\pi^0)$ 
evaluated at $M^{}_{K^+\pi^-}=M^{}_{K^*(890)}$, and 
the amplitude ${\cal A}(D^0\ra K^-\pi^+\pi^0)$ evaluated 
at $M^{}_{K^-\pi^+}=M^{}_{K^*(890)}$.

\begin{center}
\includegraphics[width=80mm]{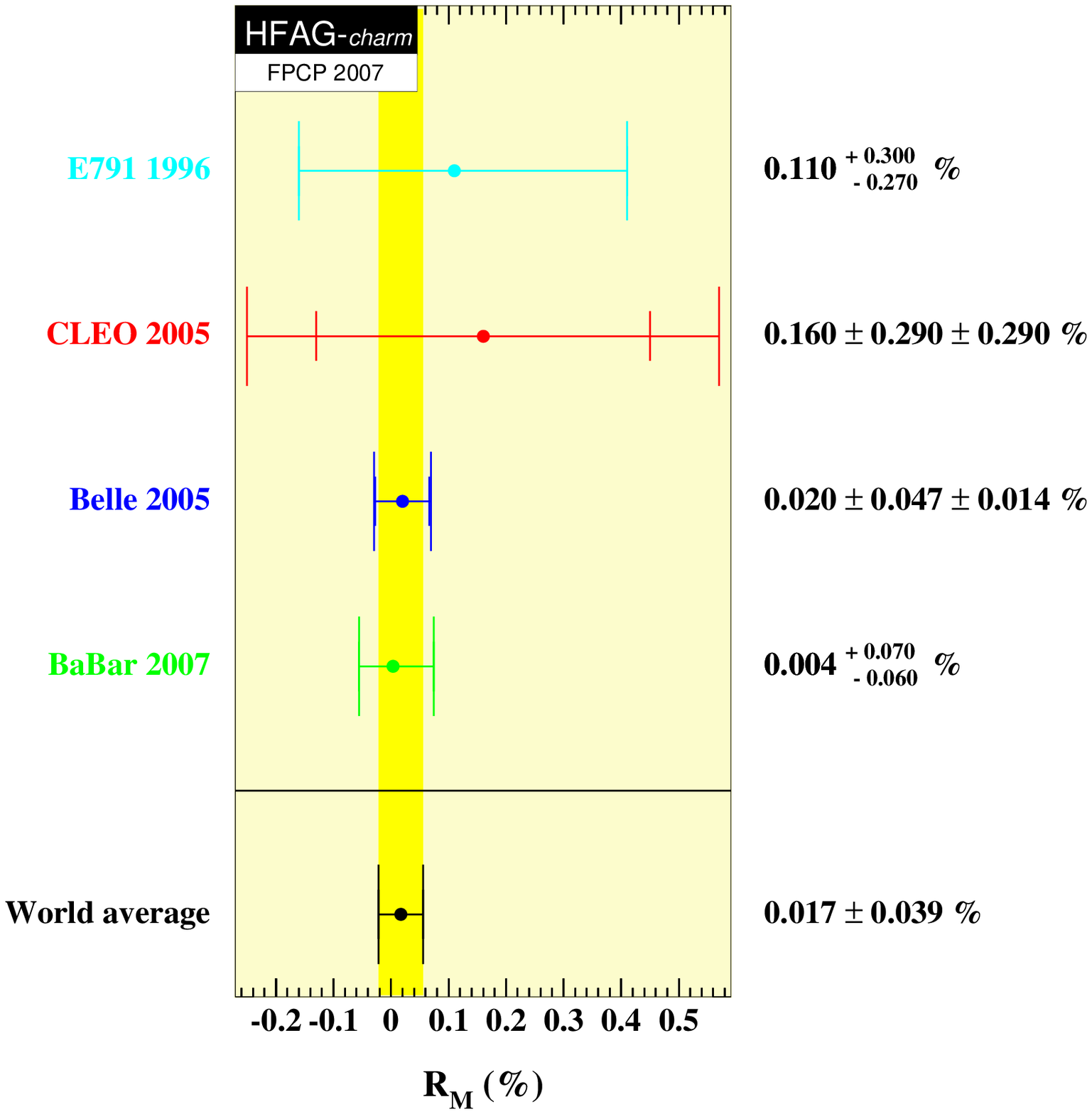}
\figcaption{\label{fig:rm_semi}
WA value of $R^{}_M$ from Ref.~\citep{hfag_charm},
as calculated from $D^0\ra K^+\ell^-\nu$ 
measurements.$^{\mbox{\tiny \citep{semi_references}}}$
}
\end{center}

All input values are listed in Table~\ref{tab:observables}. 
The observable $R^{}_M=(x^2+y^2)/2$ measured in \dklnu\ 
decays\cite{semi_references} is taken to be the WA value\cite{hfag_charm} 
calculated by HFAG (see Fig.~\ref{fig:rm_semi}). The observables 
$y^{}_{CP}$ and $A^{}_\Gamma$ measured in 
$D^0\ra K^+K^-/\pi^+\pi^-$ decays\cite{belle_kk,ycp_references} are 
also taken to be their WA values\cite{hfag_charm} (see Fig.~\ref{fig:ycp}).
The observables from \dkspp\ decays\cite{kspp_references}
for no-\cpv\ are HFAG WA values,\cite{hfag_charm} but
for the \cpv-allowed case only Belle values are available.
The \dkpi\ observables used are from Belle\cite{belle_kpi} 
and Babar,\cite{babar_kpi} as these measurements have much 
greater precision than previously published \dkpi\ results.
The $D^0\ra K^+\pi^-\pi^0$ and $D^0\ra K^+\pi^-\pi^+\pi^-$
results are from Babar,\cite{knpi_references} and the
$\psi(3770)\ra DD$ results are from CLEOc.\cite{cleoc}

The relationships between the observables and the fitted
parameters are listed in Table~\ref{tab:relationships}. 
For each set of correlated observables, we construct the 
difference vector $\vec{V}$, e.g., for 
$D^0\ra K^0_S\,\pi^+\pi^-$ decays
$\vec{V}=(\Delta x,\Delta y,\Delta |q/p|,\Delta \phi)$,
where $\Delta$ represents the difference between 
the measured value and the fitted parameter value. 
The contribution of a set of measured observables 
to the $\chi^2$ is calculated as
$\vec{V}\cdot (M^{-1})\cdot\vec{V}^T$, where
$M^{-1}$ is the inverse of the covariance matrix 
for the measurement. All covariance matrices used 
are listed in Table~\ref{tab:observables}.

\begin{center}
\includegraphics[width=80mm]{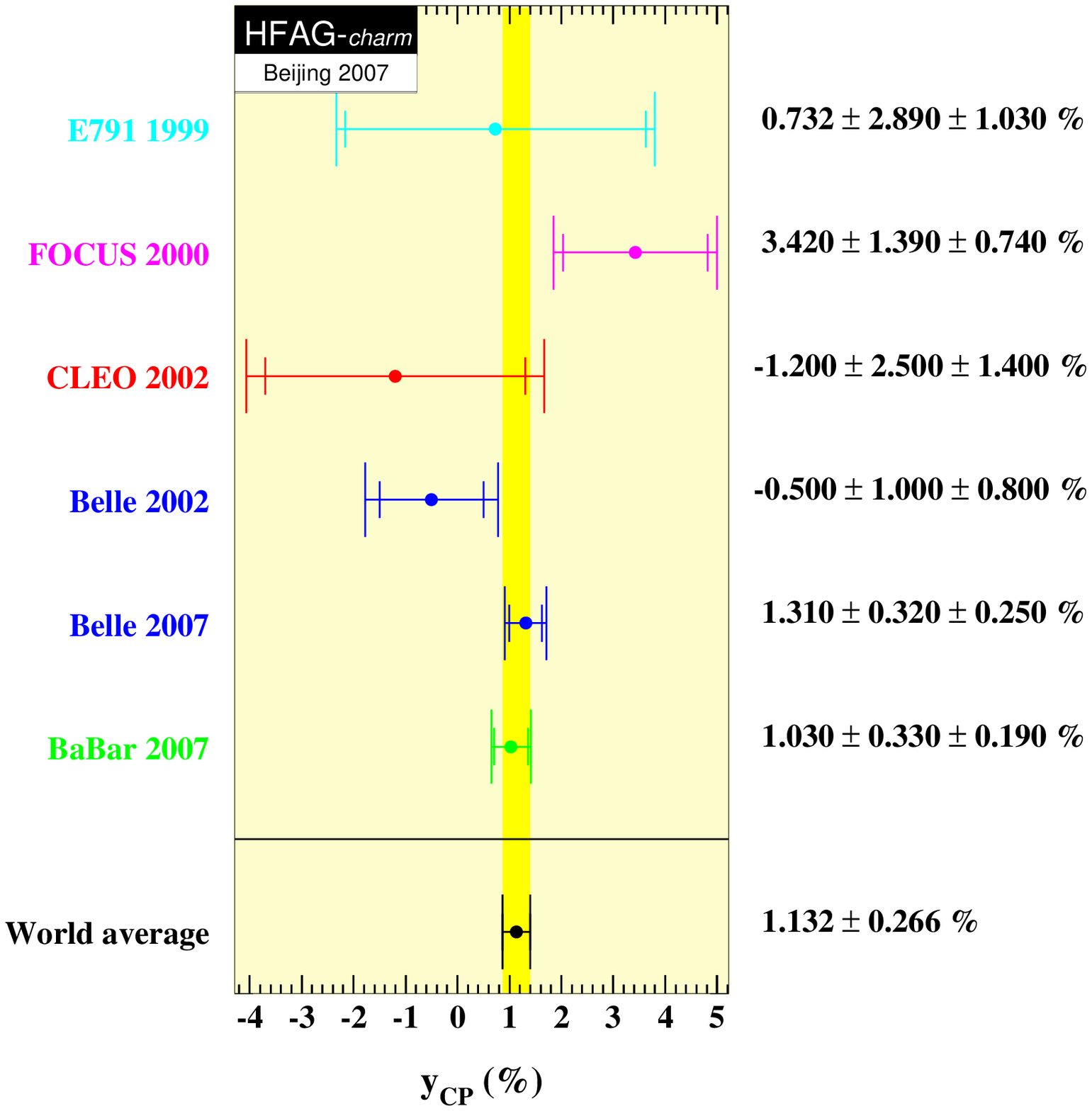}
\includegraphics[width=80mm]{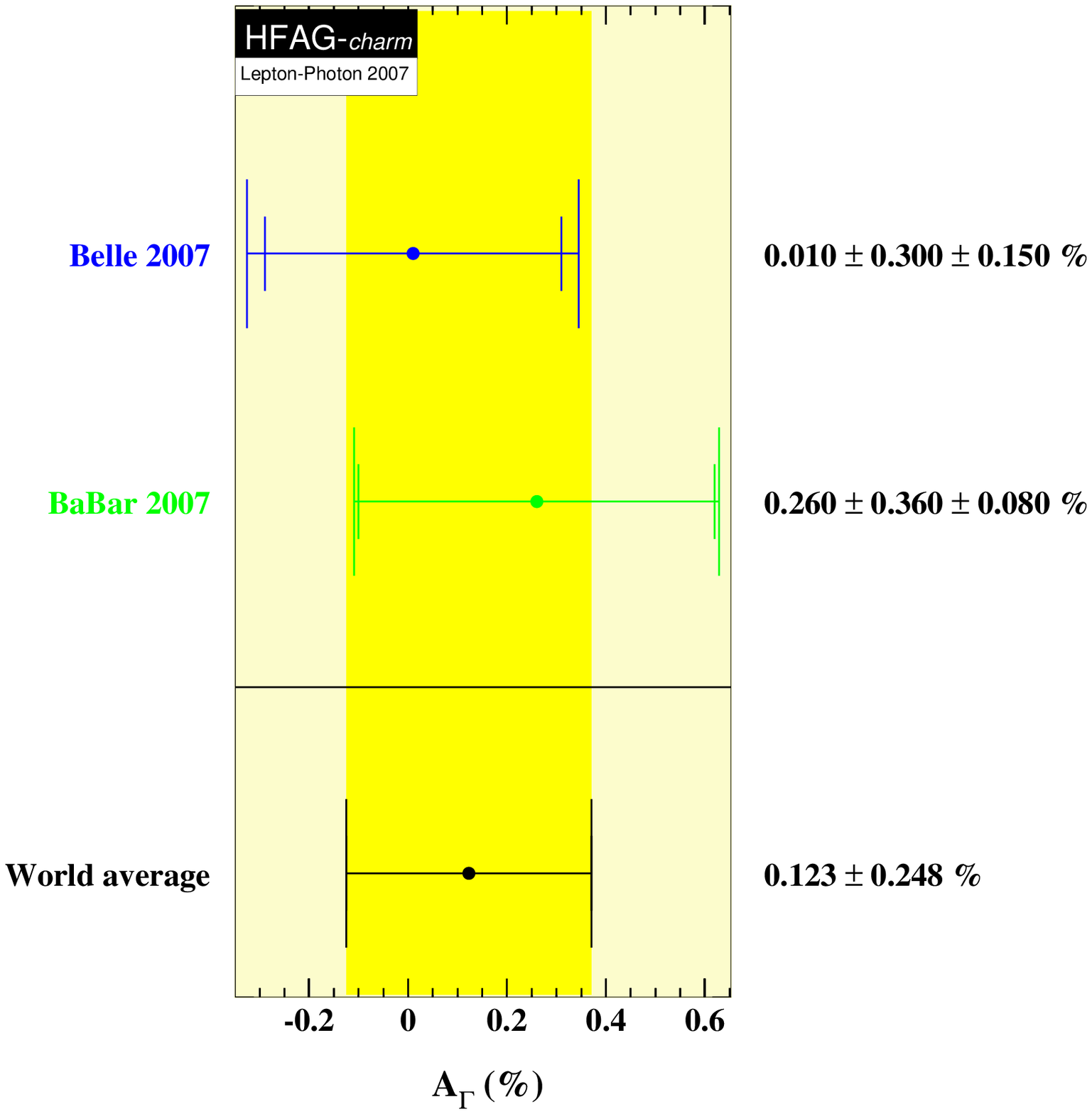}
\figcaption{\label{fig:ycp}
WA values of $y^{}_{CP}$ (top) and $A^{}_\Gamma$ (bottom)
from Ref.~\citep{hfag_charm}, as calculated from 
\dkkpp\ measurements.$^{\mbox{\tiny \citep{belle_kk,ycp_references}}}$
}
\end{center}

\end{multicols}

\begin{table}
\renewcommand{\arraystretch}{1.3}
\begin{center}
\tabcaption{\label{tab:observables}
Input values used for the global fit, from
Refs.~\citep{semi_references,belle_kk,ycp_references,
kspp_references,babar_kpi,belle_kpi,knpi_references,cleoc}.
}
\footnotesize
\begin{tabular}{ccl}
\hline\hline
\textbf{Observable} & \textbf{Value} & \textbf{Comment} \\
\hline
\begin{tabular}{c}
 $y^{}_{CP}$  \\
 $A^{}_{\Gamma}$
\end{tabular} & 
$\begin{array}{c}
(1.132\pm 0.266)\% \\
(0.123\pm 0.248)\% 
\end{array}$   &
\begin{tabular}{l}  
WA $D^0\ra K^+K^-/\pi^+\pi^-$ results~\citep{hfag_charm}
\end{tabular} \\
\hline
\begin{tabular}{c}
$x$ (no \cpv) \\
$y$ (no \cpv) \\
$|q/p|$ (no direct \cpv) \\
$\phi$ (no direct \cpv) 
\end{tabular} & 
\begin{tabular}{c}
 $(0.811\pm 0.334)\%$ \\
 $(0.309\pm 0.281)\%$ \\
 $0.95\pm 0.22^{+0.10}_{-0.09}$ \\
 $(-0.035\pm 0.19\pm 0.09)$ rad
\end{tabular} &
\begin{tabular}{l}
No \cpv: \\
WA $D^0\ra K^0_S\,\pi^+\pi^-$ results~\citep{hfag_charm}
\end{tabular} \\ 
 & & \\
\begin{tabular}{c}
$x$ \\
$y$ \\
$|q/p|$ \\
$\phi$  
\end{tabular} & 
\begin{tabular}{c}
 $(0.81\pm 0.30^{+0.13}_{-0.17})\%$ \\
 $(0.37\pm 0.25^{+0.10}_{-0.15})\%$ \\
 $0.86\pm 0.30^{+0.10}_{-0.09}$ \\
 $(-0.244\pm 0.31\pm 0.09)$ rad
\end{tabular} &
\begin{tabular}{l}
\cpv-allowed: \\
Belle $D^0\ra K^0_S\,\pi^+\pi^-$ results. Correlation coefficients: \\
\hskip0.30in $\left\{ \begin{array}{cccc}
 1 &  -0.007 & -0.255\alpha & 0.216  \\
 -0.007 &  1 & -0.019\alpha & -0.280 \\
 -0.255\alpha &  -0.019\alpha & 1 & -0.128\alpha  \\
  0.216 &  -0.280 & -0.128\alpha & 1 
\end{array} \right\}$ \\
Note: $\alpha=(|q/p|+1)^2/2$ is a variable transformation factor
\end{tabular} \\
\hline
 $R^{}_M$  & $(0.0173\pm 0.0387)\%$  &  
WA $D^0\ra K^+\ell^-\nu$ results~\citep{hfag_charm} \\
\hline
\begin{tabular}{c}
$x''$ \\ 
$y''$ 
\end{tabular} &
\begin{tabular}{c}
$(2.39\pm 0.61\pm 0.32)\%$ \\ 
$(-0.14\pm 0.60\pm 0.40)\%$ 
\end{tabular} &
\begin{tabular}{l}
Babar $D^0\ra K^+\pi^-\pi^0$ result. Correlation coefficient $=-0.34$. \\
Note: $x'' \equiv x\cos\delta^{}_{K\pi\pi} + y\sin\delta^{}_{K\pi\pi}$, 
$y'' \equiv y\cos\delta^{}_{K\pi\pi} - x\sin\delta^{}_{K\pi\pi}$.
\end{tabular} \\
\hline
 $R^{}_M$  & $(0.019\pm 0.0161)\%$  &  
Babar $D^0\ra K^+\pi^-\pi^+\pi^-$ result. \\
\hline
\begin{tabular}{c}
$R^{}_M$ \\
$y$ \\
$R^{}_D$ \\
$\sqrt{R^{}_D}\cos\delta$  
\end{tabular} & 
\begin{tabular}{c}
 $(0.199\pm 0.173\pm 0.0)\%$ \\
 $(-5.207\pm 5.571\pm 2.737)\%$ \\
 $(-2.395\pm 1.739\pm 0.938)\%$ \\
 $(8.878\pm 3.369\pm 1.579)\%$ 
\end{tabular} &
\begin{tabular}{l}
CLEOc results from ``double-tagged'' branching fractions \\
measured in $\psi(3770)\ra DD$ decays. Correlation coefficients: \\
\hskip0.30in $\left\{ \begin{array}{cccc}
 1 &  -0.0644 &  0.0072 &  0.0607 \\
 -0.0644 & 1 & -0.3172 & -0.8331 \\
 0.0072 & -0.3172 & 1 & 0.3893 \\
 0.0607 & -0.8331 & 0.3893 & 1 
\end{array} \right\}$ \\
Note: the only external input to these fit results are \\
branching fractions.
\end{tabular} \\
\hline
\begin{tabular}{c}
$R^{}_D$ \\
$x'^{2+}$ \\
$y'^+$ 
\end{tabular} & 
\begin{tabular}{c}
 $(0.303\pm 0.0189)\%$ \\
 $(-0.024\pm 0.052)\%$ \\
 $(0.98\pm 0.78)\%$ 
\end{tabular} &
\begin{tabular}{l}
Babar $D^0\ra K^+\pi^-$ results. Correlation coefficients: \\
\hskip0.30in $\left\{ \begin{array}{ccc}
 1 &  0.77 &  -0.87 \\
0.77 & 1 & -0.94 \\
-0.87 & -0.94 & 1 
\end{array} \right\}$
\end{tabular} \\
\hline
\begin{tabular}{c}
$A^{}_D$ \\
$x'^{2-}$ \\
$y'^-$ 
\end{tabular} & 
\begin{tabular}{c}
 $(-2.1\pm 5.4)\%$ \\
 $(-0.020\pm 0.050)\%$ \\
 $(0.96\pm 0.75)\%$ 
\end{tabular} &
\begin{tabular}{l}
Babar $D^0\ra K^+\pi^-$ results; correlation coefficients same as above.
\end{tabular} \\
\hline
\begin{tabular}{c}
$R^{}_D$ \\
$x'^{2+}$ \\
$y'^+$ 
\end{tabular} & 
\begin{tabular}{c}
 $(0.364\pm 0.018)\%$ \\
 $(0.032\pm 0.037)\%$ \\
 $(-0.12\pm 0.58)\%$ 
\end{tabular} &
\begin{tabular}{l}
Belle $D^0\ra K^+\pi^-$ results. Correlation coefficients: \\
\hskip0.30in $\left\{ \begin{array}{ccc}
 1 &  0.655 &  -0.834 \\
0.655 & 1 & -0.909 \\
-0.834 & -0.909 & 1 
\end{array} \right\}$
\end{tabular} \\
\hline
\begin{tabular}{c}
$A^{}_D$ \\
$x'^{2-}$ \\
$y'^-$ 
\end{tabular} & 
\begin{tabular}{c}
 $(2.3\pm 4.7)\%$ \\
 $(0.006\pm 0.034)\%$ \\
 $(0.20\pm 0.54)\%$ 
\end{tabular} &
\begin{tabular}{l}
Belle $D^0\ra K^+\pi^-$ results; correlation coefficients same as above.
\end{tabular} \\
\hline\hline
\end{tabular}
\end{center}
\end{table}

\begin{multicols}{2}

\end{multicols}

\begin{table}
\renewcommand{\arraystretch}{1.3}
\begin{center}
\tabcaption{\label{tab:relationships}
Left: decay modes used to determine fitted parameters 
$x,\,y,\,\delta,\,\delta^{}_{K\pi\pi},\,R^{}_D,\,A^{}_D,\,|q/p|$, and $\phi$.
Middle: the observables measured for each decay mode. Right: the 
relationships between the observables measured and the fitted parameters.}
\footnotesize
\begin{tabular}{l|c|l}
\hline\hline
\textbf{Decay Mode} & \textbf{Observables} & \textbf{Relationship} \\
\hline
$D^0\ra K^+K^-/\pi^+\pi^-$  & 
\begin{tabular}{c}
 $y^{}_{CP}$  \\
 $A^{}_{\Gamma}$
\end{tabular} & 
$\begin{array}{c}
2y^{}_{CP} = 
\left(\left|q/p\right|+\left|p/q\right|\right)y\cos\phi\ -\ 
\left(\left|q/p\right|-\left|p/q\right|\right)x\sin\phi \\
2A^{}_\Gamma = 
\left(\left|q/p\right|-\left|p/q\right|\right)y\cos\phi\ -\ 
\left(\left|q/p\right|+\left|p/q\right|\right)x\sin\phi
\end{array}$   \\
\hline
$D^0\ra K^0_S\,\pi^+\pi^-$ & 
$\begin{array}{c}
x \\ 
y \\ 
|q/p| \\ 
\phi
\end{array}$ &   \\ 
\hline
$D^0\ra K^+\ell^-\nu$ & $R^{}_M$  & $R^{}_M = (x^2 + y^2)/2$ \\
\hline
\begin{tabular}{l}
$D^0\ra K^+\pi^-\pi^0$ \\
(Dalitz plot analysis)
\end{tabular} & 
$\begin{array}{c}
x'' \\ 
y''
\end{array}$ &
$\begin{array}{l}
x'' = x\cos\delta^{}_{K\pi\pi} + y\sin\delta^{}_{K\pi\pi} \\ 
y'' = y\cos\delta^{}_{K\pi\pi} - x\sin\delta^{}_{K\pi\pi}
\end{array}$ \\
\hline
$D^0\ra K^+\pi^-\pi^+\pi^-$ & $R^{}_M$  & $R^{}_M = (x^2 + y^2)/2$ \\
\hline
\begin{tabular}{l}
``Double-tagged'' branching fractions \\
measured in $\psi(3770)\ra DD$ decays
\end{tabular} & 
$\begin{array}{c}
R^{}_M \\
y \\
R^{}_D \\
\sqrt{R^{}_D}\cos\delta
\end{array}$ &   $R^{}_M = (x^2 + y^2)/2$ \\
\hline
$D^0\ra K^+\pi^-$ &
$\begin{array}{c}
R^+_D,\ R^-_D \\
x'^{2+},\ x'^{2-} \\
y'^+,\ y'^-
\end{array}$ & 
$\begin{array}{l}
R^{}_D = (R^+_D + R^-_D)/2 \\
A^{}_D = (R^+_D - R^-_D)/(R^+_D + R^-_D)  \\ \\
x' = x\cos\delta + y\sin\delta \\ 
y' = y\cos\delta - x\sin\delta \\
A^{}_M\equiv (|q/p|^4-1)/(|q/p|^4+1) \\
x'^\pm = [(1\pm A^{}_M)/(1\mp A^{}_M)]^{1/4}(x'\cos\phi\pm y'\sin\phi) \\
y'^\pm = [(1\pm A^{}_M)/(1\mp A^{}_M)]^{1/4}(y'\cos\phi\mp x'\sin\phi) \\
\end{array}$ \\
\hline\hline
\end{tabular}
\end{center}
\end{table}

\begin{multicols}{2}

\section{\boldmath Fit results}

The global fit uses MINUIT with the MIGRAD minimizer, and
all errors are obtained from MINOS. Three separate fits are 
performed: {\it (a)}\ assuming \cp\ conservation ($A^{}_D$ 
and $\phi$ are fixed to zero, $|q/p|$ is fixed to one);
{\it (b)}\ assuming no direct \cpv\ ($A^{}_D$ is 
fixed to zero); and
{\it (c)}\ allowing full \cpv\ (all parameters
floated). The results are listed in 
Table~\ref{tab:results}. For the \cpv-allowed fit,
individual contributions to the $\chi^2$ are listed 
in Table~\ref{tab:results_chi2}. The total $\chi^2$ 
is 23.5 for $26-8=18$ degrees of freedom; this 
corresponds to a confidence level of~0.17.

Confidence contours in the two dimensions $(x,y)$ or 
in $(|q/p|,\phi)$ are obtained by letting, for any point in the
two-dimensional plane, all other fitted parameters take their 
preferred values. The resulting $1\sigma$-$5\sigma$ contours 
are shown in Fig.~\ref{fig:contours_ncpv} for the \cp-conserving
case, and in Fig.~\ref{fig:contours_cpv} for the \cpv-allowed 
case. The contours are determined from the increase of the
$\chi^2$ above the minimum value.
One observes that the $(x,y)$ contours for no-\cpv\ and 
for \cpv-allowed are almost identical. In both cases the $\chi^2$ 
at the no-mixing point $(x,y)\!=\!(0,0)$ is 49 units above the 
minimum value; this has a confidence level corresponding to 
$6.7\sigma$. Thus, no mixing is excluded at this high level.
In the $(|q/p|,\phi)$ plot, the point $(1,0)$ is on the 
boundary of the $1\sigma$ contour; thus the data is 
consistent with no~\cpv.

One-dimensional confidence curves for individual parameters 
are obtained by letting, for any value of the parameter, all other 
fitted parameters take their preferred values. The resulting
functions $\Delta\chi^2=\chi^2-\chi^2_{\rm min}$ (where $\chi^2_{\rm min}$
is the minimum value) are shown in Fig.~\ref{fig:1dlikelihood}.
The points where $\Delta\chi^2=2.70$ determine 90\% C.L. intervals 
for the parameters as shown in the figure. The points where 
$\Delta\chi^2=3.84$ determine 95\% C.L. intervals; these are
listed in Table~\ref{tab:results}.

\begin{center}
\includegraphics[width=80mm]{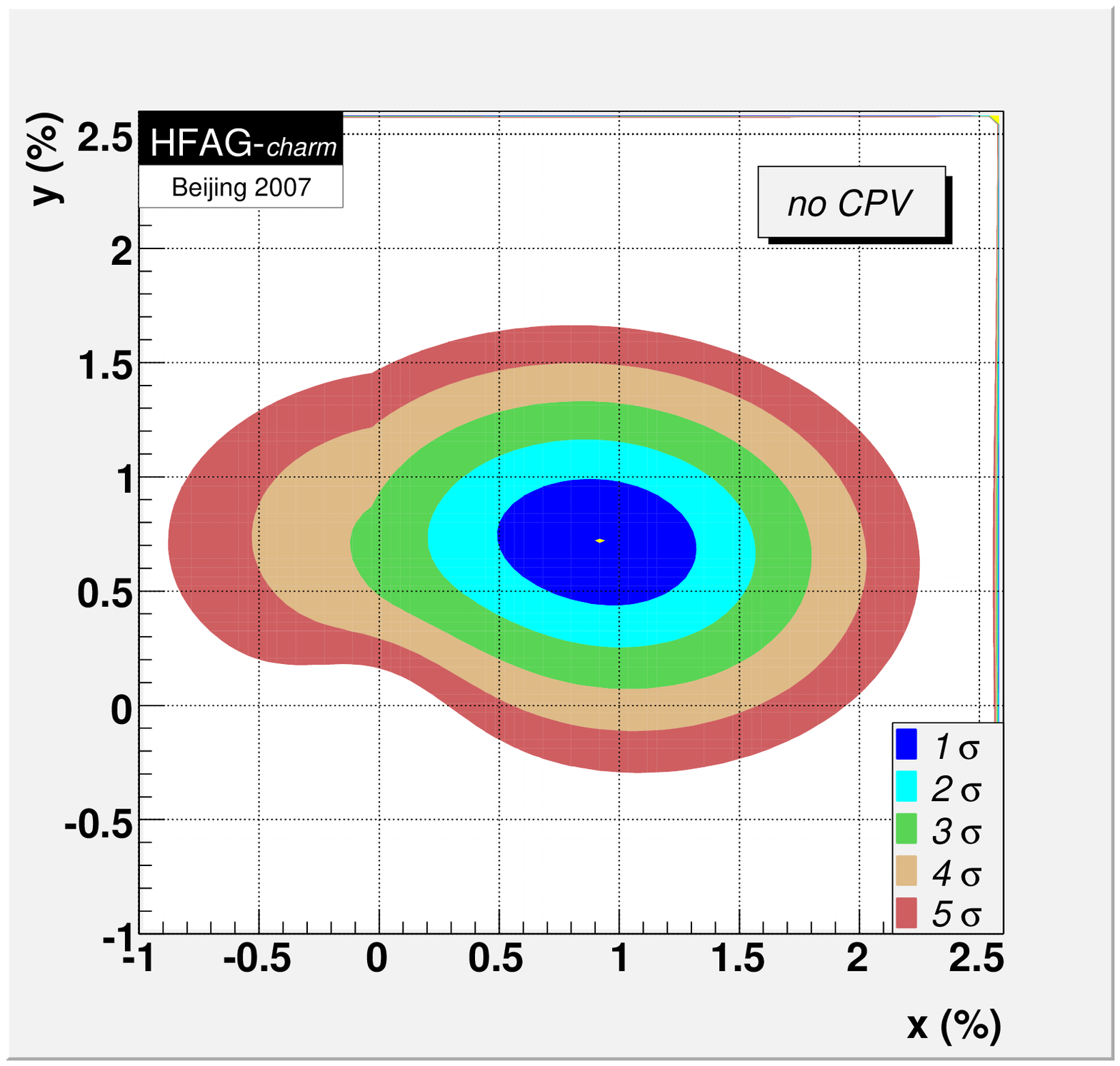}
\figcaption{\label{fig:contours_ncpv}
Two-dimensional contours for mixing parameters $(x,y)$, for no \cpv. }
\end{center}

\end{multicols}

\begin{center}
\hbox{
\includegraphics[width=80mm]{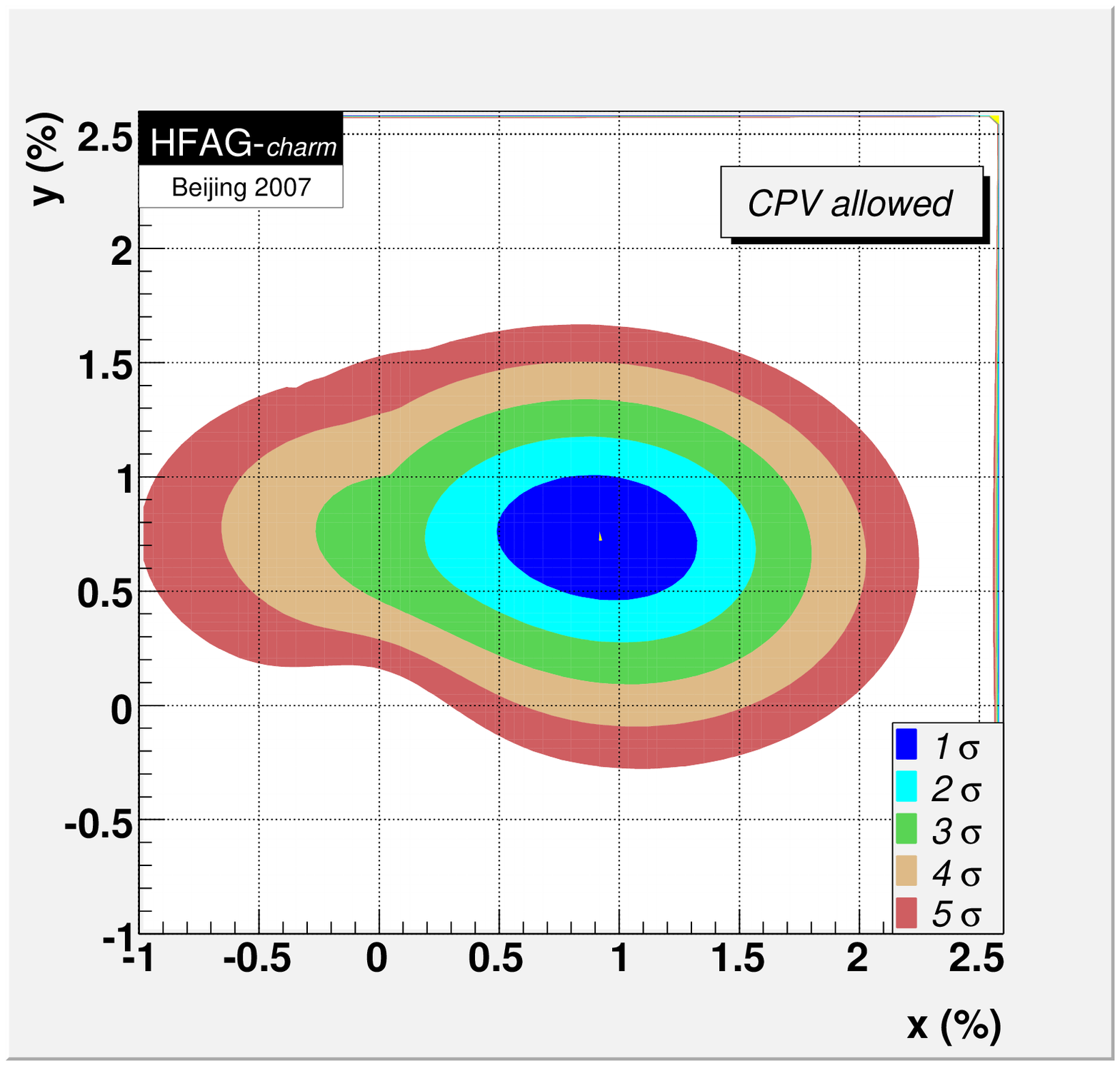}
\hskip0.15in
\includegraphics[width=80mm]{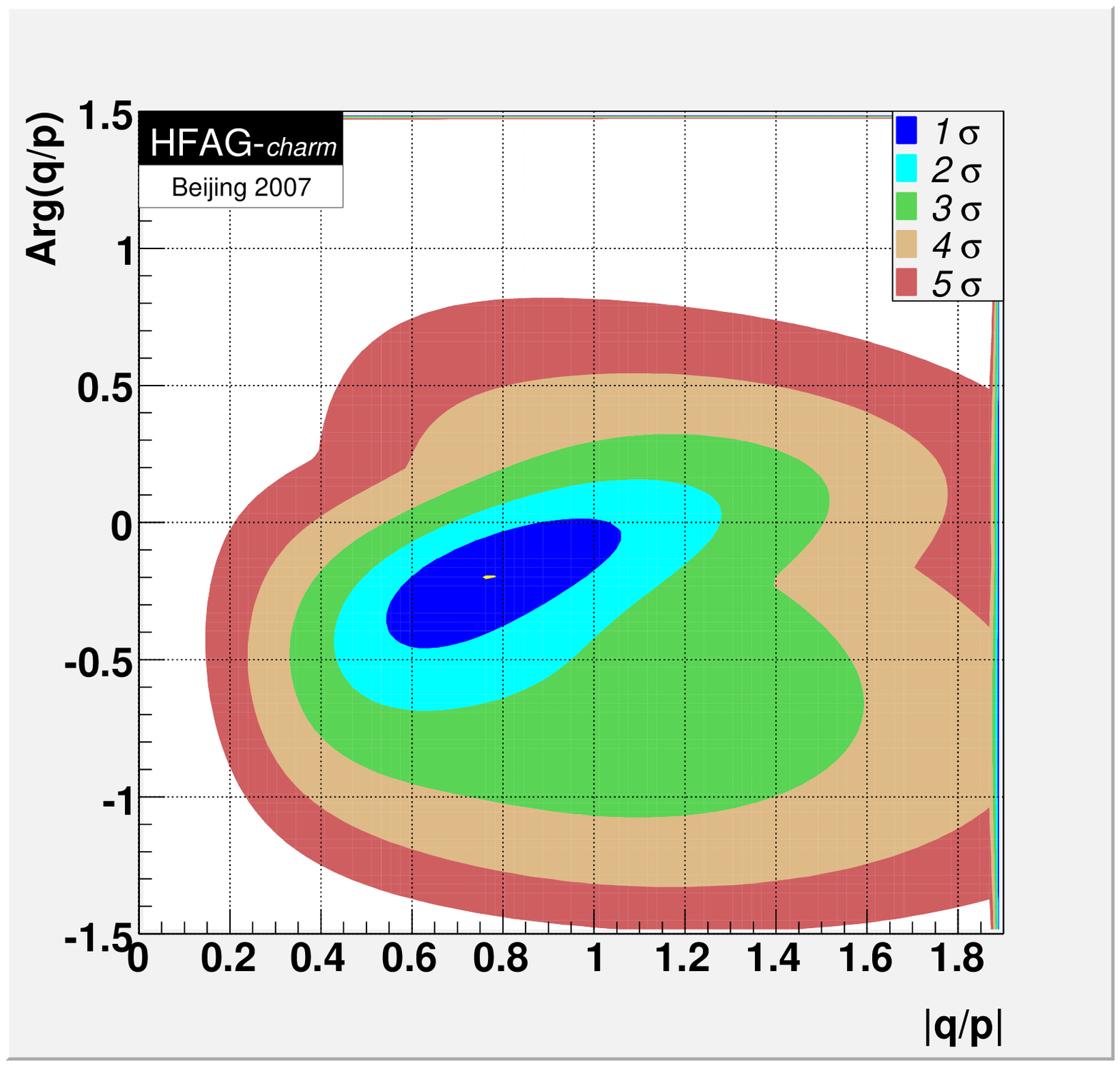}
}
\figcaption{\label{fig:contours_cpv}
Two-dimensional contours for parameters $(x,y)$ (left) 
and $(|q/p|,\phi)$ (right), allowing for \cpv.}
\end{center}

\begin{center}
\hbox{\hskip0.50in
\includegraphics[width=72mm]{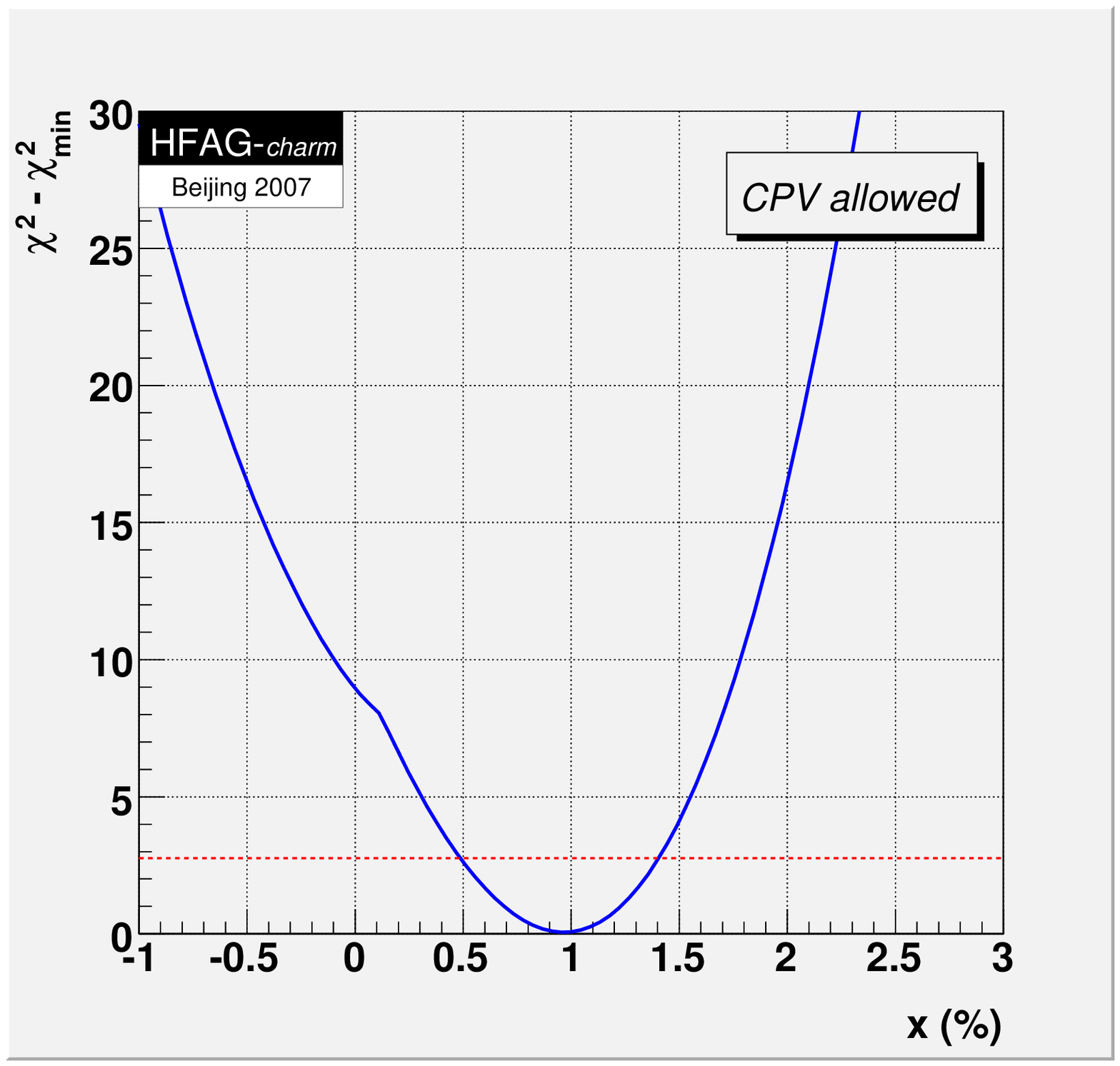}
\hskip0.20in
\includegraphics[width=72mm]{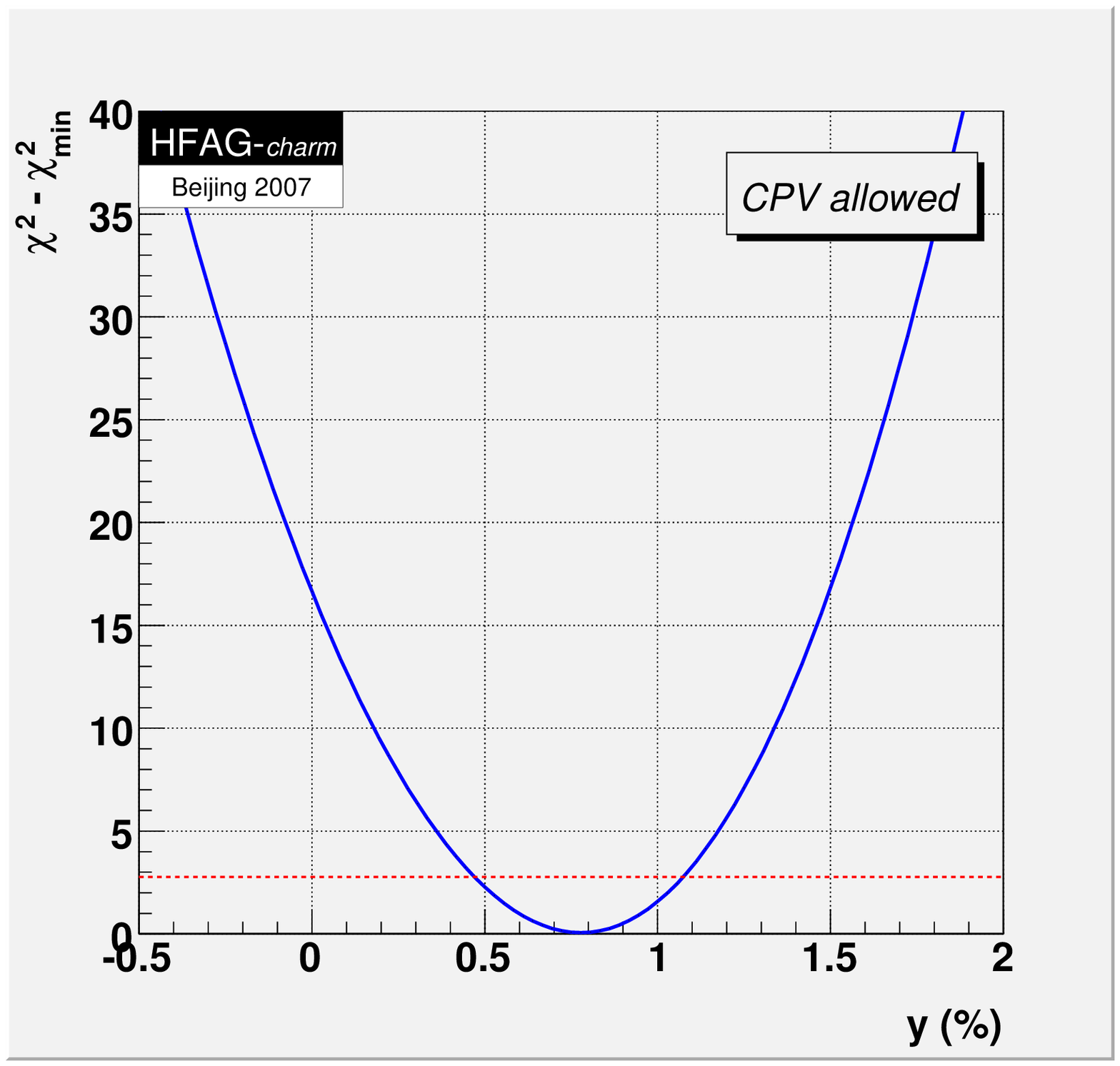}}
\hbox{\hskip0.50in
\includegraphics[width=72mm]{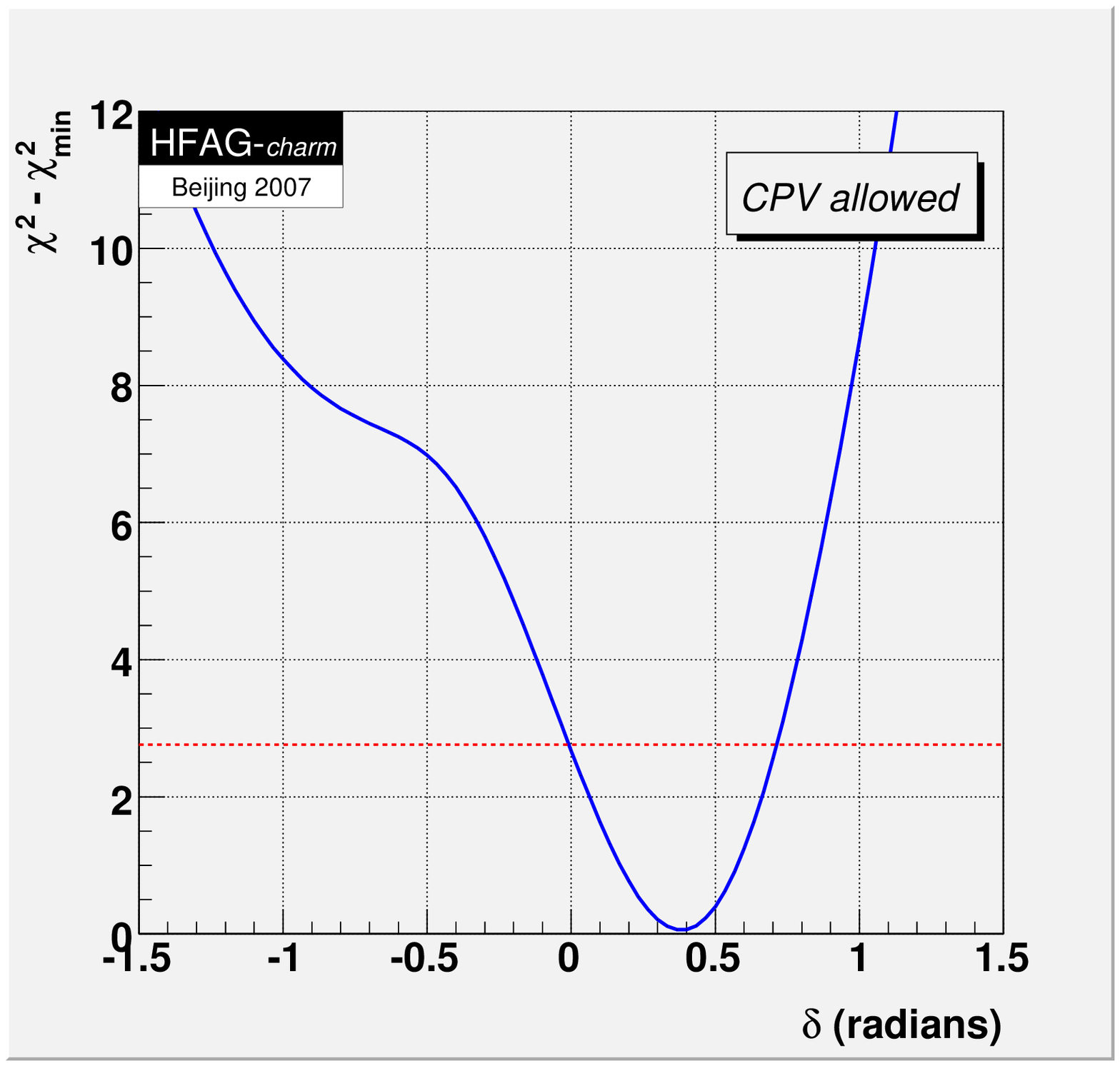}
\hskip0.20in
\includegraphics[width=72mm]{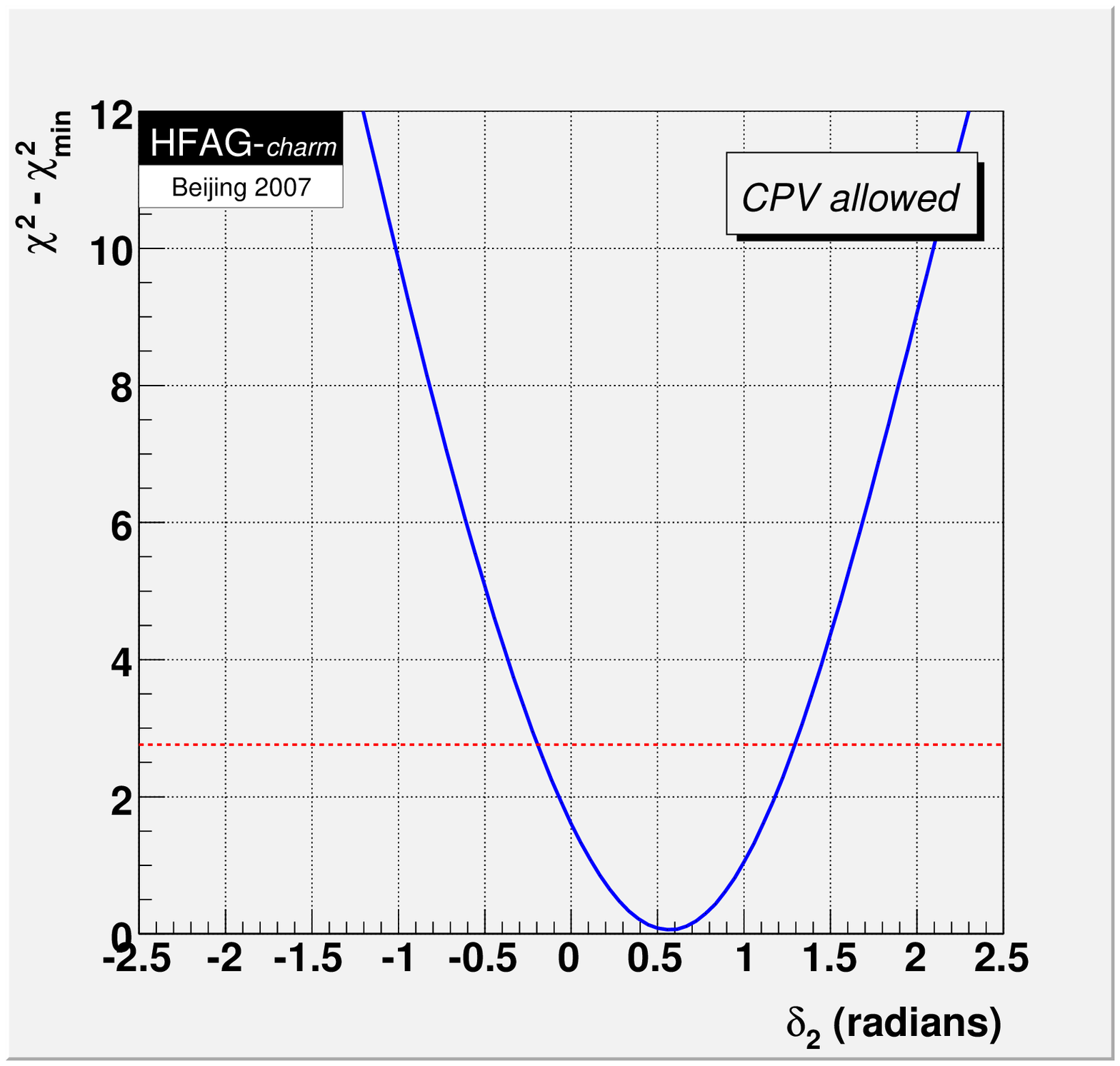}}
\hbox{\hskip0.50in
\includegraphics[width=72mm]{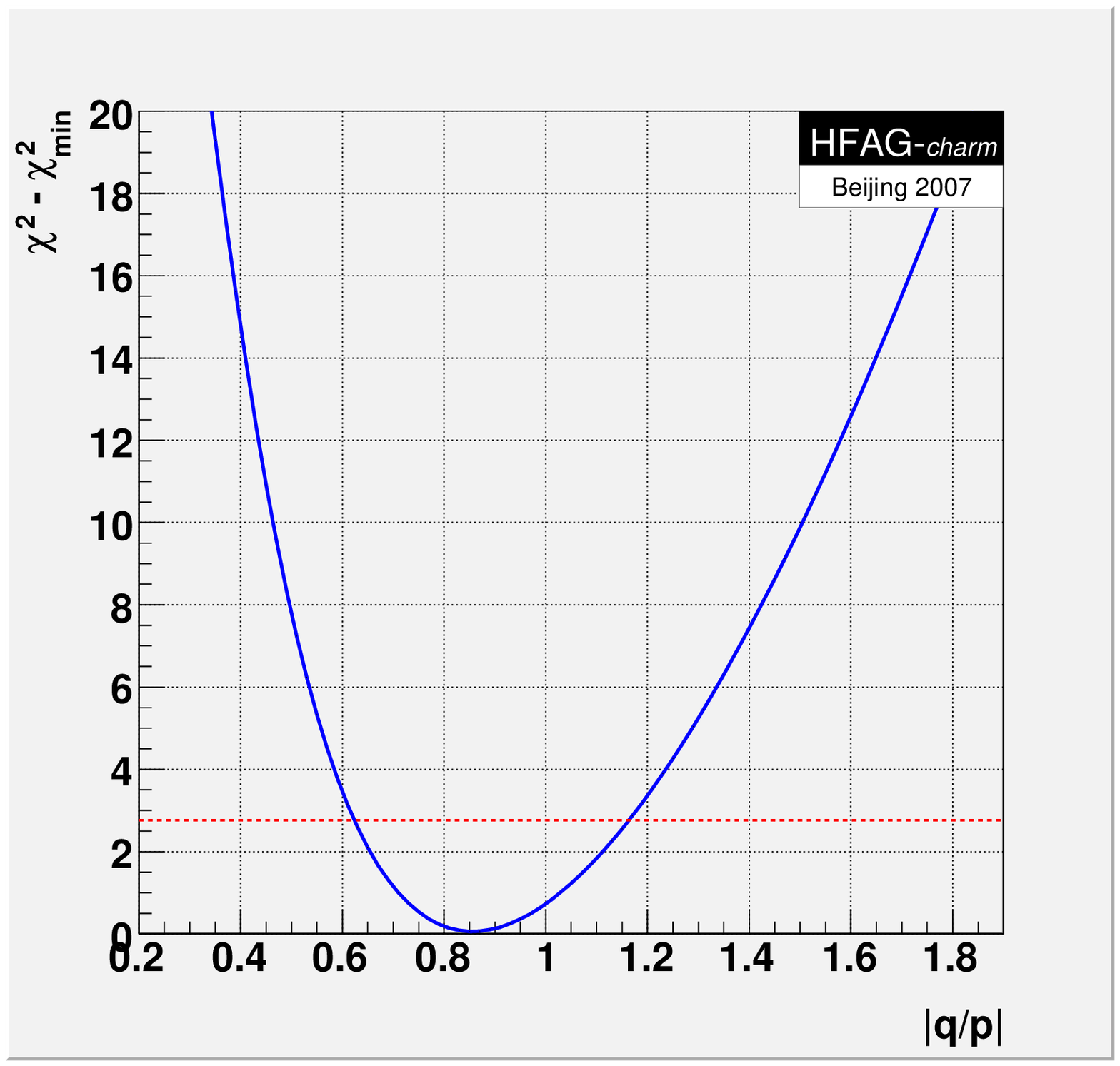}
\hskip0.20in
\includegraphics[width=72mm]{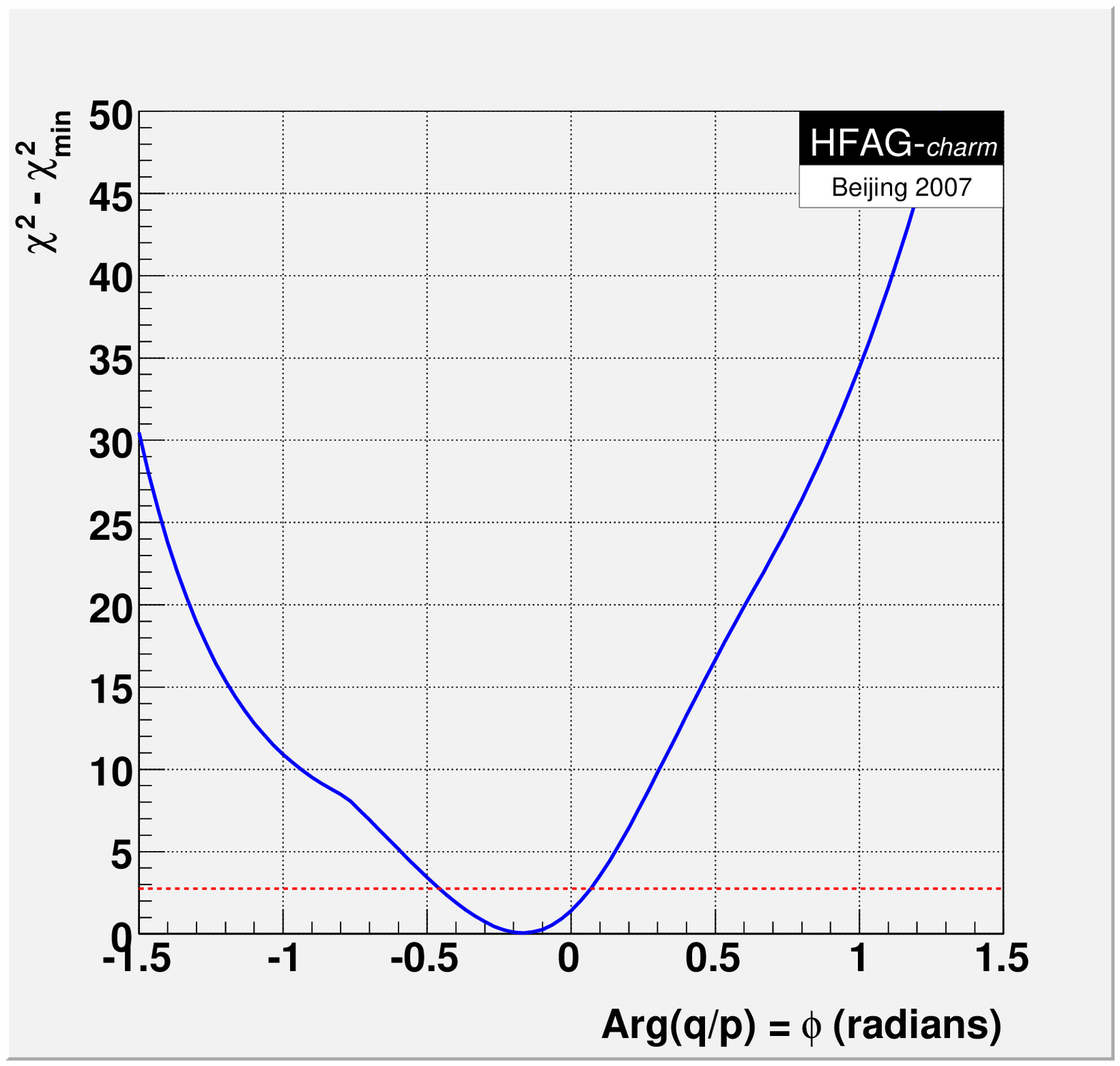}}
\vskip0.12in
\figcaption{\label{fig:1dlikelihood}
The function $\Delta\chi^2=\chi^2-\chi^2_{\rm min}$ 
for fitted parameters
$x,\,y,\,\delta,\,\delta^{}_{K\pi\pi},\,|q/p|$, and $\phi$.
The points where $\Delta\chi^2=2.70$ (denoted by the dashed 
horizontal line) determine a 90\% C.L. interval. }
\end{center}

\begin{table}
\renewcommand{\arraystretch}{1.4}
\begin{center}
\tabcaption{\label{tab:results}
Results of the global fit for different assumptions concerning~\cpv.}
\footnotesize
\begin{tabular}{c|cccc}
\hline\hline
\textbf{Parameter} & \textbf{\boldmath No \cpv} & \textbf{\boldmath No direct \cpv} 
& \textbf{\boldmath \cpv-allowed} & \textbf{\boldmath \cpv-allowed 95\% C.L.}  \\
\hline
$\begin{array}{c}
x\ (\%) \\ 
y\ (\%) \\ 
\delta\ (^\circ) \\ 
R^{}_D\ (\%) \\ 
A^{}_D\ (\%) \\ 
|q/p| \\ 
\phi\ (^\circ) \\
\delta^{}_{K\pi\pi}\ (^\circ)  
\end{array}$ & 
$\begin{array}{c}
0.98\,^{+0.26}_{-0.27} \\
0.75\,\pm 0.18 \\
21.6\,^{+11.6}_{-12.6} \\
0.335\,\pm 0.009 \\
- \\
- \\
- \\
30.8\,^{+25.0}_{-25.8} 
\end{array}$ &
$\begin{array}{c}
0.97\,^{+0.27}_{-0.29} \\
0.78\,^{+0.18}_{-0.19} \\
23.4\,^{+11.6}_{-12.5} \\
0.334\,\pm 0.009 \\
- \\
0.95\,^{+0.15}_{-0.14} \\
-2.7\,^{+5.4}_{-5.8} \\ 
32.5\,^{+25.0}_{-25.7} 
\end{array}$ &
$\begin{array}{c}
0.97\,^{+0.27}_{-0.29} \\
0.78\,^{+0.18}_{-0.19} \\
21.9\,^{+11.5}_{-12.5} \\
0.335\,\pm 0.009 \\
-2.2\,\pm 2.5 \\
0.86\,^{+0.18}_{-0.15} \\
-9.6\,^{+8.3}_{-9.5} \\ 
32.4\,^{+25.1}_{-25.8} 
\end{array}$ &
$\begin{array}{c}
 0.39 - 1.48 \\
 0.41 - 1.13 \\
-6.3 - 44.6 \\
0.316 - 0.353 \\
-7.10 - 2.67 \\
 0.59 - 1.23 \\
-30.3 - 6.5 \\
-20.3 - 82.7 
\end{array}$ \\
\hline\hline
\end{tabular}
\end{center}
\end{table}

\begin{table}
\renewcommand{\arraystretch}{1.4}
\begin{center}
\tabcaption{\label{tab:results_chi2}
Individual contributions to the $\chi^2$ for the \cpv-allowed fit.}
\footnotesize
\begin{tabular}{l|rr}
\hline\hline
\textbf{Observable} & \textbf{\boldmath $\chi^2$} & \textbf{\boldmath $\sum\chi^2$} \\
\hline
$y^{}_{CP}$ & 2.06 & 2.06 \\
$A^{}_\Gamma$ & 0.10 & 2.16 \\
\hline
$x^{}_{K^0\pi^+\pi^-}$ & 0.20 & 2.36 \\
$y^{}_{K^0\pi^+\pi^-}$ & 1.94 & 4.30 \\
$|q/p|^{}_{K^0\pi^+\pi^-}$ & 0.00 & 4.30 \\
$\phi^{}_{K^0\pi^+\pi^-}$ & 0.46 & 4.76 \\
\hline
$R^{}_M(K^+\ell^-\nu)$ & 0.06 & 4.83 \\
\hline
$x^{}_{K^+\pi^-\pi^0}$ & 1.24 & 6.06 \\
$y^{}_{K^+\pi^-\pi^0}$ & 1.62 & 7.69 \\
\hline
$R^{}_M/y/R^{}_D/\sqrt{R^{}_D}\cos\delta$ (CLEOc) & 5.59 & 13.28 \\
\hline
$R^+_D/x'{}^{2+}/y'{}^+$ (Babar) & 2.54 & 15.82 \\
$R^-_D/x'{}^{2-}/y'{}^-$ (Babar) & 1.75 & 17.57 \\
$R^+_D/x'{}^{2+}/y'{}^+$ (Belle) & 3.96 & 21.53 \\
$R^-_D/x'{}^{2-}/y'{}^-$ (Belle) & 1.43 & 22.95 \\
\hline
$R^{}_M(K^+\pi^-\pi^+\pi^-)$ & 0.49 & 23.45 \\
\hline\hline
\end{tabular}
\end{center}
\end{table}

\newpage
\begin{multicols}{2}

\section{\boldmath Conclusions}

From the global fit results listed in Table~\ref{tab:results}
and shown in Figs.~\ref{fig:contours_cpv} and \ref{fig:1dlikelihood},
we conclude the following:
\begin{itemize}
\item the experimental data consistently indicate that 
$D^0$ mesons undergo mixing. The no-mixing point $x=y=0$
is excluded at $6.7\sigma$. The parameter $x$ differs from
zero by $3.0\sigma$; the parameter $y$ differs from zero by
$4.1\sigma$. The effect is presumably dominated 
by long-distance processes, which are difficult to calculate.
Thus unless $|x|\gg |y|$ (see Ref.~\citep{BigiUraltsev}), 
it may be difficult to identify new physics from mixing alone.
\item Since \ycp\ is positive, the \cp-even state is shorter-lived,
as in the $K^0$-$\kbar$ system. However, since $x$ also appears
to be positive, the \cp-even state is heavier, 
unlike in the $K^0$-$\kbar$ system.
\item It appears difficult to accomodate a strong phase
difference $\delta$ larger than $45^\circ$.
\item There is no evidence yet for \cpv\ in the $D^0$-$\dbar$ system.
Observing \cpv\ at the level of sensitivity of the current experiments 
would indicate new physics.
\end{itemize}

\vspace{2mm}

\vspace{-2mm}
\centerline{\rule{80mm}{0.1pt}}

\end{multicols}

\clearpage

\end{document}